\documentclass{WileyMSP-template}
\ifx\dedication\undefined
\usepackage[margin=1in]{geometry}
\else
\begin{document}
\pagestyle{fancy}
\fi
\title{Comment on ``Evolution Equations of Nonlinearly Permissible,
Coherent Hole Structures Propagating Persistently in
Collisionless Plasmas''}

\ifx\dedication\undefined
\date{Plasma Science and Fusion Center,\\ Massachusetts Institute of
Technology,\\ Cambridge MA 02139, USA.}
\else
\maketitle
\fi

\author{I H Hutchinson}

\ifx\dedication\undefined
\begin{document}
\maketitle
\else
\begin{affiliations}
Plasma Science and Fusion Center,\\ Massachusetts Institute of
Technology,\\ Cambridge MA 02139, USA\\
Email Address: ihutch@mit.edu
\end{affiliations}

\keywords{Electron hole, Ion hole, Vlasov equation, Poisson Equation}
\fi

\begin{abstract}
  Recent critical remarks, published in \iftrue this journal\else
  ``Annalen der Physik''\fi, about the present author's analysis of
  electron and ion holes and their stability are addressed and shown
  to be misunderstandings and misrepresentations.
\end{abstract}

\section*{Comment}

In a recent publication\cite{Schamel2023a}, Schamel and Chakrabarti,
make a point of contradicting some features of my papers concerning
electron and ion holes. While I have in many places acknowledged the
important pioneering work of Dr. Schamel concerning electron holes,
there remain various important aspects of his claims on which we
disagree. Normally I am content to present my analysis and findings
objectively in my own papers, in the expectation that the give and
take of the progress of science will eventually clarify which
perspective or opinion is most satisfactory. However, the referenced
paper misrepresents my positions so egregiously that I feel it is
essential to comment and correct the misrepresentations.

In my 2017 tutorial paper about electron holes\cite{Hutchinson2017}, I
tried, for the benefit of the community, to explain the decades-long
dispute (in which I had no part) between what Schamel calls
respectively the ``pseudo-potential method'' and the ``BGK
method''. (Actually Bernstein Greene and Kruskal\cite{Bernstein1957}
introduced both methods and called them the ``differential equation''
and ``integral equation'' methods.) I said, and I remain firmly
convinced, that these are both valid ways to obtain equilibrium
electron hole solutions, and they have different strengths and
weaknesses. The pseudo-potential approach specifies the trapped
particle velocity distribution function $f(v)$ and solves the steady
Vlasov-Poisson system to obtain the corresponding self-consistent
potential $\phi(x)$. Its strength is that $f(v)$ can be specified so
as to avoid some singularities; however it requires the solution of an
eigenproblem to ensure that the trapping boundary energy lies at the
assumed value in the prescribed $f(v)$; and its weakness is that
historically monotonic trapped $f(v)$ of specific form were generally
assumed, which substantially constrains its characteristics. (Recent
generalizations by Schamel of the $f(v)$ form permit, for example,
$f(v)$ discontinuities, thus weakening one of its strengths and
strengthening one of its weaknesses.) The BGK method, in contrast,
specifies the potential profile and solves for the required trapped
distribution function. Its strength is that seemingly any potential
shape is permitted, and that no eigensolution is required; its
weakness is that unless certain plausibility constraints are applied
to $\phi(x)$, notably that it falls off asymptotically exponentially
with the Debye shielding length, then slope singularities in $f(v)$
arise at the trapping boundary (and possibly at other potential
extrema).

In \cite{Schamel2023a} Schamel and Chakrabarti say concerning
my comparison of the two methods that ``Hutchinson ... does not
adequately acknowledge their differences. In particular, he sees no
need for an independent derivation of a NDR''. It is true that I see
no need for an eigensolution in the BGK method if one prescribes the
asymptotic form of $\phi$ correctly. It is not true that the BGK
method requires ``independent derivation of a NDR''; instead the
asymptotic constraint becomes what Schamel calls the NDR (Nonlinear
Dispersion Relation). Nor is it true that in every case ``it would be
better to start directly with the pseudo-potential method''. Both
methods have advantages, but for different purposes. I also disagree
with their abstract's extreme claim that ``structure formation is
inevitably associated with particle trapping, which can only be
properly addressed by the pseudo-potential method''. Trapping and
electron hole formation is generally highly dynamic and non-linear,
often involving merging of smaller holes into larger ones; and neither
equilibrium construction method can meaningfully represent the
formation processes.

They further say ``he keeps making mistakes in his calculations'' with
the only justification being a reference to a longer
version\cite{Schamel2023} of \cite{Schamel2023a}. (The authors call
\cite{Schamel2023a} an ``abridged version'' of \cite{Schamel2023}). In
the longer version, the only passage that appears to refer to my
``mistakes'' is that in a footnote of my tutorial
paper\cite{Hutchinson2017} I remark on, and explain in sufficient
detail that others can check, a
n algebraic error I thought I had detected in a Taylor expansion in a
paper by Korn and Schamel \cite{Korn1996}. On rechecking that
observation, I find that, though there was an error in their
development where I said, their subsequent equations (A4 on) and final
expressions were actually correct. I therefore withdraw the remark
with apologies to Korn and Schamel, and acknowledge that the final
coefficient in equation (12) of \cite{Hutchinson2017} should be $1/16$
as they maintain, not $1/32$ as I wrote.  Fortunately, as I said,
``these errors are not of great significance''.  

The difference in our perspectives is starkest when Schamel and
Chakrabarti dispute the findings of \cite{Hutchinson2023} where I
explain that there is a sign error in the original paper of Schamel
and Bujabarua 1980, in their application to ion hole temperature ratio
requirements of what Schamel calls the NDR, and what I see (in the BGK
approach) as the constraint on the asymptotic form of $\phi(x)$
(effectively the same equation). To come to some resolution of this
important issue, which has often been cited, perhaps a more direct
comparison of our hole treatments might be helpful.

My version of the asymptotic constraint (i.e.\ the distant boundary
condition) when employing the BGK method is simply Poisson's equation
\begin{equation}
  (1/q_r) d^2\phi/dx^2=n_{f}+\tilde n -n_{r},
\end{equation}
where $q_r=\pm1$ is the charge (sign) and $n_r$ is the density of the
repelled species (ions for a positive potential); $n_f$ is the
attracted species density including a flat trapped $f(v)$ equal to the
passing value $f({\cal E}=0)$; and $\tilde n$ is the (usually
negative) deviation of the actual trapped density caused by the
deviation of trapped $f(v)$ from flat: effectively the phase space
density hole. Asymptotically, the potential decays exponentially to
zero at large $|x|$: $\phi \propto \exp(-|x|/\lambda)$; so
$d^2\phi/dx^2=\phi/\lambda^2$. Moreover
$n_{r}\simeq 1+(dn_{r}/d\phi) \phi$, and
$n_{f}\simeq 1+(dn_{f}/d\phi) \phi$, while if $f(v)$ is continuous
with finite derivatives at the trapping boundary ${\cal E}=0$, then
$\tilde n \sim \phi^{3/2}$. Thus in the limit $\phi\to 0$ it is
required that $1/\lambda^2=1/\lambda_{Da}^2+1/\lambda_{Dr}^2$ where
$1/\lambda_{D}^2=\lim_{\phi\to0}|dn/d\phi|$, and $\lambda_D$ is
the Debye length of each species (for general velocity distribution
functions). This requirement is nothing other than familiar Debye
shielding of potential in a plasma. But it invokes continuity of
attracted species $f(v)$ at ${\cal E}=0$ to justify ignoring the then
higher order $\tilde n$. Note that $\theta=T_r/T_a$ is the ratio of
the repelled species temperature to the attracted temperature. For an
ion hole this is $T_e/T_i$.

Schamel's version of the NDR is most easily related to mine from the
longer version paper \cite{Schamel2023}, to whose equations and
notation I now refer. When his parameters $k_0$, $\tilde C$, $D_1$ and
$D_2$ are zero we have ``The privileged ${\rm sech}^4(x)$ solitary
mode'', eq (12). Its potential form is
$\phi(x) =\psi {\rm\, sech}^4(\sqrt{B} x/4)$, and gives a NDR eq (13)
\begin{equation}
  -{1\over2}Z'_r(\tilde v_D/\sqrt{2}) -{\theta\over2}Z'_r(u_0/\sqrt{2}) =
  B - \Gamma.
\end{equation}
Here the $-Z'_r$ function terms (subscript $r$ denotes real part in
his notation) represent
$\lim_{\phi\to 0}|dn/d\phi|={1/\lambda_D^2}$ for the two species,
which are Maxwellians with shifts $\tilde v_D$ and $u_0$; and
evidently from the potential form, $B$ is equal to the asymptotic
inverse potential scale length squared, my $1/\lambda^2$. Thus if
$\Gamma=0$, the equation is identical to my asymptotic constraint. But
$\Gamma\;(\equiv{\sqrt{\pi}\over 2}\exp(-v_D^2/2)\gamma)$ represents
(see \cite{Schamel2023a}, eq 1) a constant added into the trapped
distribution function but not to the passing distribution. Therefore,
my requirement of continuity of $f(v)$ across ${\cal E}=0$ amounts to
the assumption that $\Gamma=0$.

By contrast, in their arguments to defend the temperature ratio
condition that I have contradicted, Schamel and Chakrabarti say ``We
neglect for simplicity in (1) the right hand side and get
$k_0^2-{1\over2}Z'_r(v_0/\sqrt{2}) -{\theta\over2}Z'_r(u_0/\sqrt{2})
=0$'' and then ``For ion acoustic waves ... using
$-{1\over2}Z'_r(u_0/\sqrt{2})\approx 0$ one gets
$-{1\over2}Z'_r(u_0/\sqrt{2})=-1/\theta(k_0^2+1)$. This is essentially
(14) of his paper...''. This statement is incorrect even when $k_0=0$,
the solitary limit. Actually, my eq (13) (not 14, which has nothing to
do with the issue) is the one to compare; and it has \emph{not}
neglected all the right hand terms that in Schamel's notation are
represented by $B$. Instead, it recognizes that the
$B\equiv 1/\lambda^2$ term for a solitary potential must be positive,
and it supposes the repelled species is an unshifted Maxwellian,
giving $-{1\over2}Z'_r(\bar v_a/\sqrt{2})=1/\lambda^2-1/\theta$ (where
$\bar v_a$ is the Maxwellian shift of the attracted species). From
this follows trivially that when
$-{1\over2}Z'_r(\bar v_a/\sqrt{2})=-0.285$ (its global minimum over
all $\bar v_a$) it is required that $\theta < 1/0.285=3.5$ as I have
stated, not $\theta > 3.5$ as Schamel continues to maintain based on
improperly neglecting the important right hand side term(s). Others
before me. e.g.\ \cite{Chen2004}, whom I cite in
\cite{Hutchinson2023}, have given counterexamples of equilibria
disproving Schamel's version.

I believe the source of the confusion lies in the reference to ion
acoustic waves, which do of course require $\theta \gg 1$ for low
damping. An ion hole, however, is not like an ion acoustic soliton,
nor a nonlinear extension of ion acoustic waves, both of which can be
treated using a fluid (single speed) ion representation. An ion hole
is intrinsically a kinetic phenomenon involving a depression in phase
space density (the hole) within the spread of the ion velocity
distribution. So their statement ``And of course the same
holds for ion holes with $u_0 = 1.307$'' is an incorrect assumption,
based on a mistaken analogy with ion acoustic phenomena.
It is ironic that Schamel and Chakrabarti later say of me ``In
summary, many of his ideas are still influenced by linear wave theory,
which, however, is misplaced in this area.'' In fact, they are the ones
misled by analogies with ion acoustic waves. I, by contrast, have
consistently treated holes as requiring full-scale kinetic treatment
of the attracted species, as is their nature.

A final misrepresentation I must contradict consists in Schamel and
Chakraborti's remark saying that in my work `invalid results such as
“ultra- slow velocities exist only at the center of a double-humped
ion distribution” are obtained.' In none of my published papers can I
find the passage they place in double quotes implying that it is a
quotation. The nearest thing I can find to a statement along these
lines is that in \cite{Hutchinson2021c} I write ``for slow positive
structures to exist stably, the background ion velocity distribution
generally cannot be single-humped.'' However the crucial point is the
word ``stably''. What my analysis shows is that in the absence of a
local distribution minimum, slow electron holes are unstable to
self-acceleration. The entire point of \cite{Hutchinson2021c} and
\cite{Hutchinson2022} is stability analysis, not merely equilibrium
analysis. Slow hole lifetimes when unstable are relatively short,
because they become fast.

I hope the present comment might help the community and Schamel and
Chakraborti, to understand better the relationship between their work
and mine, and become more accepting of the value of different
perspectives, and of constructive criticism.

\medskip
\textbf{Acknowledgements} \par 
This work was not supported by any external funding. The author
reports no conflict of interest. No data were produced or used.

\bibliographystyle{MSP}
\bibliography{JabRef}

\begin{thebibliography}{1}
\providecommand{\url}[1]{\texttt{#1}}
\providecommand{\urlprefix}{URL }

\bibitem{Schamel2023a}
H.~Schamel, N.~Chakrabarti,
\newblock \emph{Annalen der Physik} \textbf{2023}, 2300102.

\bibitem{Hutchinson2017}
I.~H. Hutchinson,
\newblock \emph{Physics of Plasmas} \textbf{2017}, \emph{24}, 5 055601.

\bibitem{Bernstein1957}
I.~B. Bernstein, J.~M. Greene, M.~D. Kruskal,
\newblock \emph{Physical Review} \textbf{1957}, \emph{108}, 4 546.

\bibitem{Schamel2023}
H.~Schamel,
\newblock \emph{Reviews of Modern Plasma Physics} \textbf{2023}, \emph{7}, 1.

\bibitem{Korn1996}
J.~Korn, H.~Schamel,
\newblock \emph{Journal of Plasma Physics} \textbf{1996}, \emph{56}, 02 307.

\bibitem{Hutchinson2023}
I.~H. Hutchinson,
\newblock \emph{Physics of Plasmas} \textbf{2023}, \emph{30}, 3 032107.

\bibitem{Chen2004}
L.-J. Chen, D.~Thouless, J.-M. Tang,
\newblock \emph{Physical Review E} \textbf{2004}, \emph{69}, 5 55401.

\bibitem{Hutchinson2021c}
I.~H. Hutchinson,
\newblock \emph{Phys. Rev. E} \textbf{2021}, \emph{104} 015208.

\bibitem{Hutchinson2022}
I.~Hutchinson,
\newblock \emph{Journal of Plasma Physics} \textbf{2022}, \emph{88}, 1
  555880101.

\end{thebibliography}


\end{document}